# ORBIT SIMULATION FOR THE DETERMINATION OF RELATIVISTIC AND SOLAR-SYSTEM PARAMETERS FOR THE ASTROD SPACE MISSION

Dah-Wei Chiou[1] and Wei-Tou Ni[1]

[1]*Center for Gravitation and Cosmology,*
*Department of Physics, National Tsing Hua University,*
*Hsinchu, Taiwan 30055, ROC*

## ABSTRACT

ASTROD (Astrodynamical Space Test of Relativity using Optical Devices) mission concept is to conduct high-precision measurement of relativistic effects, solar-system parameters and gravitational waves. In a previous paper, we have presented the results of orbit simulation using a stochastic model to calculate the accuracy of determining the relativistic parameters $\beta$ and $\gamma$, and the solar quadrupole parameter $J_2$ for a simple implementation using two spacecraft plus earth system reference. In this paper, we first extend the stochastic model to simulate the determination of the masses of three big asteroids (Ceres, Vesta and Pallas). With one range observation per day for each spacecraft from 25 days to 800 days of the mission and ten range observations per day for each spacecraft from 800 days to 1050 days of the mission (when the apparent positions of the two spacecraft are close to the Sun), the accuracies of determining these parameters are $4.6 \times 10^{-7}$ for $\gamma$, $4.0 \times 10^{-7}$ for $\beta$, $1.2 \times 10^{-8}$ for $J_2$, and $6.4 \times 10^{-5} M_{Ceres}$, $7.6 \times 10^{-4} M_{Pallas}$, $8.1 \times 10^{-5} M_{Vesta}$ for the mass determination of Ceres, Pallas and Vesta respectively. We then include in the simulation and determination the rate of change of the gravitational constant ($\dot{G}$), and an anomalous constant acceleration ($a_a$) directed towards the Sun (proposed by Anderson *et al.* [*PRL* **81**, 2858(1998)] in their analysis of the radio metric data from the Pioneer 10/11, Galileo, and Ulysses spacecraft). At the end of 1050 days of simulation, the accuracies of determining these two parameters are $9.5 \times 10^{-15}$/yr for $\dot{G}/G$ and $2.0 \times 10^{-16}$ m/s$^2$ for $a_a$. The mass loss rate of the inner solar system is at the level of $1 \times 10^{-13} M_{Sun}$/yr. The accurate determination of $\dot{G}/G$ will mean a need to simultaneously determine the mass loss rate. In other words, the mission gives a way to monitor the mass loss rate of the Sun if $\dot{G}/G$ can be determined independently.

## INTRODUCTION

Lunar laser ranging experiment achieves an accuracy of 1-2 cm. It tests relativistic gravity with similar accuracy as solar-system radio tests in various aspects (Dickey *et al.*, 1994; Williams *et al.*, 1996). It measures earth system parameters (including lunar parameters) to high precision (Dickey *et al.*, 1994). With this success, active solar system ranging has been proposed. Both pulse ranging---SORT (Melliti *et al.*, 2000) & IPLR, (Smith, 2000), and interferometric ranging---Mini-ASTRO, ASTROD, Super-ASTROD (Bec-Borsenberger *et al.*, 2000) have been considered. In 1993, we have proposed to use laser astrodynamics to study relativistic gravity and to explore the solar system (Ni, 1993). With a multi-purpose astrodynamical mission proposed in 1994 (Ni, Wu and Shy, 1996; Ni *et al.*, 1997), we reached the ASTROD (Astrodynamical Space Test of Relativity using Optical Devices) mission concept.

The objectives of ASTROD mission are threefold. The first objective is to discover and explore fundamental physical laws governing matter, space and time via testing relativistic gravity with 3-6 orders of magnitude improvement. The second objective of the ASTROD mission is the high-precision measurement of the solar-system parameters. The third objective is to detect and observe gravitational waves from massive black holes and galactic binary stars in the frequency range 50 µHz to 5 mHz. Background gravitational-waves will also be explored. In January, 2000, we formed an international team and submitted the ASTROD proposal to ESA (Bec-Borsenberger *et al.*, 2000). A desirable implementation is to have two spacecraft in separate solar orbits each carrying a payload of a proof mass, two telescopes, two 1-2 W lasers, a clock and a drag-free system, together with an Earth reference system. The Earth reference system could be ground stations, Earth satellites and/or spacecraft near Earth-Sun Lagrange points. For technological development, please see Bec-Borsenberger *et al.* (2000) and references therein.

In a previous paper, we started orbit simulation and parameter determination for this simple implementation (Chiou and Ni, 2000). We use a stochastic model to calculate the accuracy of determining the relativistic parameters $\beta$ and $\gamma$, and the solar quadrupole parameter $J_2$. In this paper, we extend the stochastic model to simulate the determination of the masses of three big asteroids (Ceres, Vesta and Pallas), the rate of change of the gravitational constant, an anomalous constant acceleration directed towards the Sun proposed by Anderson *et al.* (1998) in their analysis of the radio metric data from the Pioneer 10/11, Galileo, and Ulysses spacecraft.

In the following section, we formulate the CGC 1 ephemeris and determine the perturbations on the ASTROD spacecraft. In Section 3, we generate our stochastic model and use Kalman filtering to calculate the accuracies of parameters. At the end, we conclude with a discussion.

## CGC 1 EPHEMERIS AND ORBITS OF THE ASTROD SPACECRAFT

For ASTROD orbit simulation, we need a working ephemeris to include parameters and effects under study. In our previous paper (Chiou and Ni, 2000), we have built a working ephemeris to study the determination of the relativistic parameters $\beta$ and $\gamma$, and the solar quadrupole parameter $J_2$. We used the post-Newtonian barycentric metric with solar quadrupole moment and the associated equations of motion to generate computer-integrated ephemeris for nine-planets, the Moon and the Sun. The positions and velocities at the fiducial starting point of the mission, 2005.6.10 0:00, are taken from the DE 403 ephemeris (Standish *et al.*, 1995). The evolution is solved by the $4^{\text{th}}$-order Runge-Kutta method with the stepsize h = 0.01 day.

Here, we include the 3 big asteroids --- Ceres, Pallas and Vesta to extend the 11 body evolution to 14 body evolution. We call this extended ephemeris the CGC 1 ephemeris (CGC: Center for Gravitation and Cosmology). The initial positions and velocities of the 3 big asteroids are calculated from MPO98 (1997). For the masses, we use the DE 403 values.

To include the effects of $\dot{G}$ and $a_a$, we extend the deterministic model for evaluation of partials. First, we replace the gravitational constant G by $[G_0 + \dot{G} \times (t - t_0)]$ in the post-Newtonian equations of motion to find the partials for $\dot{G}$ at different epochs. $G_0$ is the value of G at the initial epoch $t_0$. Since the post-Newtonian terms are small compared with the Newtonian terms, the $\dot{G}$ terms in the post-Newtonian terms are negligible and can be dropped out. In 1998, Anderson *et al.* proposed that the radio metric data from the Pioneer 10/11, Galileo, and Ulysses spacecraft indicated an apparent anomalous, constant, acceleration acting on the spacecraft with a magnitude of $a_a \sim 8.5 \times 10^{-8}$ cm/s$^2$, directed towards the Sun. To include this effect in the parameter-fitting, we add an acceleration term $a_a \mathbf{r}/r$ to the equation of motion. $a_a$ is the anomalous acceleration parameter to be determined.

The fiducial orbits of the two ASTROD spacecraft are calculated with $\gamma = \beta = 1$, $J_2 = 2 \times 10^{-7}$, $\dot{G} = a_a = 0$, $M_{\text{Ceres}} = 4.64 \times 10^{-10} M_{\text{Sun}}$, $M_{\text{Pallas}} = 1.05 \times 10^{-10} M_{\text{Sun}}$, and $M_{\text{Vesta}} = 1.34 \times 10^{-10} M_{\text{Sun}}$. The initial conditions are the same as in Chiou and Ni (2000). The orbits of inner spacecraft and outer spacecraft are designed such that they complete 3 rounds and 2 rounds respectively in 2.5 years (about 900 days). The initial conditions are especially chosen that the two spacecraft are near the Sun in apparent position from 800 days to 1050 days after launch --- within w $3\tau$ of the Sun from 850 to 1000 days and within w $8\tau$ of the Sun from 800 to 1050 days. During this period of time, $\gamma$ parameter (Shapiro effect parameter) accuracy can be improved greatly; accompanying this, $\beta$, $J_2$ and some other parameters can also be improved significantly. The perturbations due to the 3 big asteroids on the inner and outer spacecraft are calculated. These perturbations extend to 100-200 m during 1050 days of the mission. The orbits and perturbations agree with previous results (Su *et al.*, 1999 [In the case of perturbation on the outer spacecraft, a wrong file was taken for fig 2(b) in this reference; the correct file agrees (Su *et al.*, private communication).]).

## SIMULATION

The ranging observation $Z_k$ at time instant $t_k$ is expressed as an explicit function h(u, $t_k$) of uncertain parameters plus a superposing noise term $V_k$ as $Z_k = h(u,t_k) + V_k$, where $\mathbf{u} = (u_1, u_2...)$ are uncertain parameters to be estimated, e.g., $\gamma$, $\beta$, $J_2$, $M_{\text{Ceres}}$ etc. $\mathbf{u}$ normally includes masses and initial states of planets. However, for simulating the uncertainties of relativistic parameter determination, we only need to include a few such parameters. In the noise term $V_k$, we include two types of errors:

(1) the uncertainty due to the imprecision of the ranging devices,
(2) unknown accelerations due to the imperfections of the spacecraft drag-free system.

As in our previous paper (Chiou and Ni, 2000), the first type of error is modelled as a Gaussian random noise with zero mean and with standard deviation $5 \times 10^{-11}$ sec; for the second type of error, the magnitude of the unknown acceleration is treated as a Gaussian random noise with zero mean and with standard deviation $10^{-15}$ m/s$^2$ and the direction of the unknown acceleration is changed randomly every four hours. The deviations of the simulated rangings from the fiducial ranging are similar to those in the Fig. 2 of our previous paper.

In this paper, we make two fits. In the first one, we use 24 parameters in the fitting, i.e., $\mathbf{u} = (u_1, u_2, \ldots, u_{24}) = (x_0^{\text{inner}}, y_0^{\text{inner}}, z_0^{\text{inner}}, v_{x0}^{\text{inner}}, v_{y0}^{\text{inner}}, v_{z0}^{\text{inner}}, x_0^{\text{outer}}, y_0^{\text{outer}}, z_0^{\text{outer}}, v_{x0}^{\text{outer}}, v_{y0}^{\text{outer}}, v_{z0}^{\text{outer}}, M_{\text{Sun}}, M_{\text{Mercury}}, M_{\text{Venus}}, M_{\text{Earth+Moon}}, M_{\text{Mars}}, M_{\text{Jupiter}}, M_{\text{Ceres}}, M_{\text{Pallas}}, M_{\text{Vesta}}, \gamma, \beta, J_2)$. Six parameters are for the initial position and velocity of the inner spacecraft; six for those of the outer spacecraft; six for the masses of Sun, Mercury, Venus, Earth-Moon System, Mars and Jupiter; three for the masses of Ceres, Pallas and Vesta; two for the relativistic parameters $\gamma$ and $\beta$; and one for the solar quadrupole parameter $J_2$. In the second fit, we add two parameters, $\dot{G}$ and $a_a$, to make a 26-parameter fit. In calculating the partials, we linearize the equations of motion around the fiducial trajectory with deviations small enough so that we can neglect the nonlinear errors but large enough so that numerical round-off errors are not important. In the fitting, we use Kalman sequential filtering method as in our previous paper. We take one range observation per day for each spacecraft from 25 days to 800 days of the mission and ten range observations per day for each spacecraft from 800 days to 1050 days of the mission (when the apparent positions of the two spacecraft are close to the Sun). At 25.00 day after launch, first range observation of the inner spacecraft is made; at 25.50 day, first range observation of the outer spacecraft is made. These observations alternate with inner spacecraft and outer spacecraft until 800.00 day after launch. Starting at 800.00 day, these alternations switch with a time interval of 0.05 days until 1050 day after launch.

In the first fit, the uncertainties of $\gamma$, $\beta$ and $J_2$ as functions of epoch are shown in Fig. 1; the fractional uncertainties of $M_{\text{Ceres}}$, $M_{\text{Pallas}}$, $M_{\text{Vesta}}$ and $M_{\text{Mars}}$ as functions of epoch are shown in Fig . 2. The accuracies of fitted relativistic and solar-system parameters after 1050 days of the mission are as follows:

$\gamma$,    $4.6 \times 10^{-7}$;           $M_{\text{Ceres}}$,      $6.4 \times 10^{-5} M_{\text{Ceres}}$;

$\beta$,    $4.0 \times 10^{-7}$;           $M_{\text{Pallas}}$,     $7.6 \times 10^{-4} M_{\text{Pallas}}$;

$J_2$,    $1.2 \times 10^{-8}$;           $M_{\text{Vesta}}$,      $8.1 \times 10^{-5} M_{\text{Vesta}}$

with the uncertainty of Mars mass as $2.7 \times 10^{-9} M_{\text{Mars}}$. The solar rotation axis is tilted $7 \text{r } 15'$ to the perpendicular of the ecliptic. For this paper, we take it perpendicular to the ecliptic; the error due to this simplification is consistent with the reached $J_2$ uncertainty. In the future, we will include the tilt angle in our ephemeris.

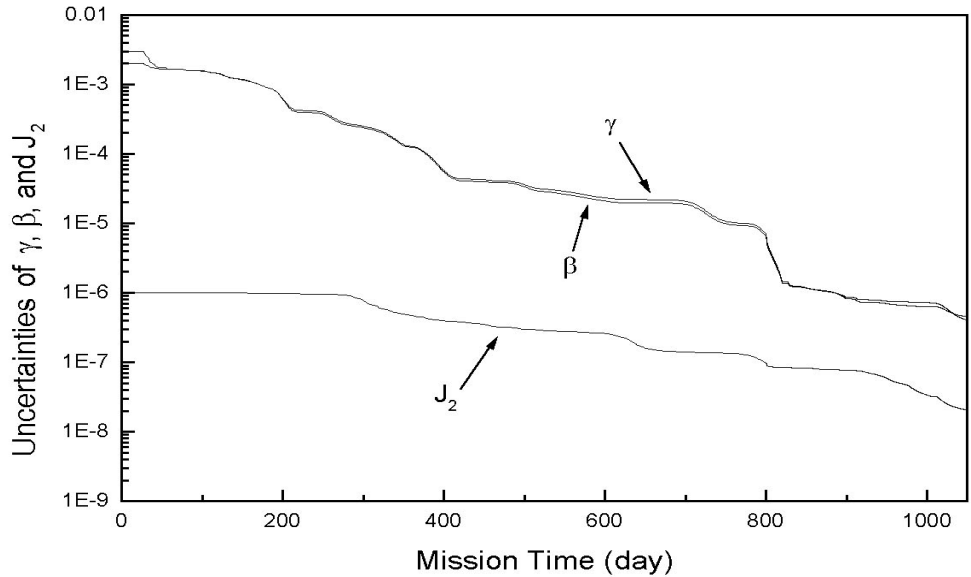

Fig. 1. Uncertainties of $\gamma$, $\beta$ and $J_2$ as functions of epoch.

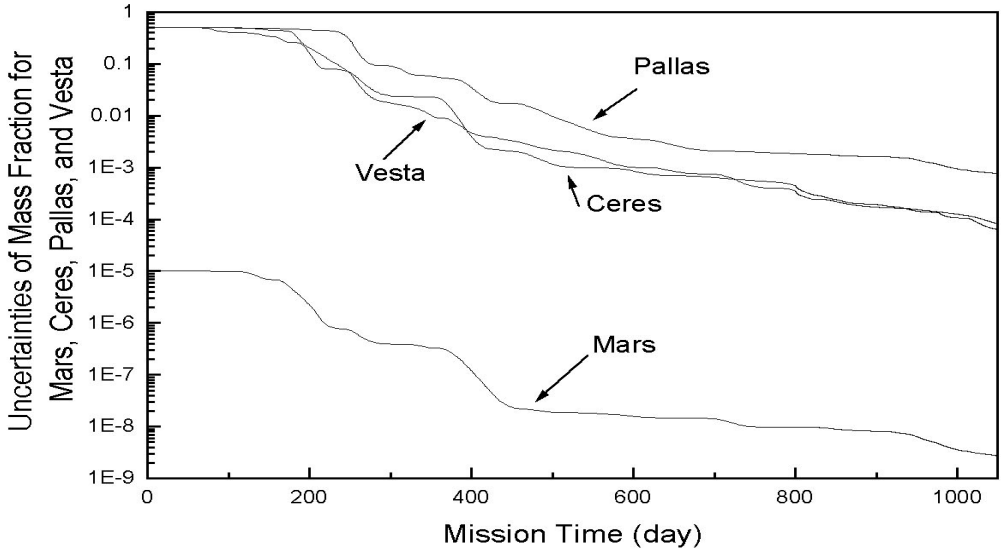

Fig. 2. Fractional uncertainties of $M_{Ceres}$, $M_{Pallas}$, $M_{Vesta}$ and $M_{Mars}$ as functions of epoch.

In the estimation of relativistic parameters of our previous paper (Chiou and Ni, 2000), we use 15 parameters in the fitting, i.e., $\mathbf{u} = (u_1, u_2, \ldots, u_{15}) = (x_0^{inner}, y_0^{inner}, z_0^{inner}, v_{x0}^{inner}, v_{y0}^{inner}, v_{z0}^{inner}, M_{Sun}, M_{Mercury}, M_{Venus}, M_{Earth+Moon}, M_{Mars}, M_{Jupiter}, \gamma, \beta, J_2)$, and only use the inner spacecraft ranging data in the fitting simulation. The results are shown in Fig. 3 of that paper. After 1050 days of the mission, the accuracies of $\gamma$, $\beta$ and $J_2$ are determined to be

$\gamma$,      5.8 X $10^{-7}$;

$\beta$,      7.3 X $10^{-7}$;

$J_2$,      4.0 X $10^{-8}$.

Comparing with the above results, we see that the fitting with the ranging data of both spacecraft enhances the accuracies by a factor of 1.3-3.3. (The accuracies mentioned in the paragraph preceding Fig. 3 of our previous paper are slightly different due to shorter ending periods.)

In the second fit, we add two parameters, $\dot{G}$ and $a_a$, to make a 26-parameter fit, i.e., $\mathbf{u} = (u_1, u_2, \ldots, u_{26}) = (x_0^{inner}, y_0^{inner}, z_0^{inner}, v_{x0}^{inner}, v_{y0}^{inner}, v_{z0}^{inner}, x_0^{outer}, y_0^{outer}, z_0^{outer}, v_{x0}^{outer}, v_{y0}^{outer}, v_{z0}^{outer}, M_{Sun}, M_{Mercury}, M_{Venus}, M_{Earth+Moon}, M_{Mars}, M_{Jupiter}, M_{Ceres}, M_{Pallas}, M_{Vesta}, \gamma, \beta, J_2, \dot{G}, a_a)$. In this fit, the uncertainty of $\dot{G}$ as a function of epoch is shown in Fig. 3; the uncertainty of $a_a$ as a function of epoch is shown in Fig. 4. The uncertainty of $\dot{G}$ after 1050 days of the mission is $9.5 \times 10^{-15}$ /yr; the uncertainty of $a_a$ is $2.0 \times 10^{-16}$ m/s$^2$; the uncertainties of other parameters change only slightly from those of the first fit. The mass loss rate of the inner solar system is at the level of $1 \times 10^{-13}$ $M_{Sun}$/yr. The expected mass loss rates of the Sun due to solar electromagnetic radiations, solar winds, solar neutrinos and solar axions are respectively $7 \times 10^{-14}$ $M_{Sun}$ /yr, $\sim 10^{-14}$ $M_{Sun}$ /yr, $\sim 2 \times 10^{-15}$ $M_{Sun}$ /yr and $\sim 10^{-15}$ $M_{Sun}$ /yr (Ni, 1996). The accurate determination of $\dot{G}/G$ will mean a need to simultaneously determine the mass loss rate. In other words, the ASTROD mission gives a way to monitor the mass loss rate of the Sun if $\dot{G}/G$ can be determined independently.

For a better estimation of the uncertainty in the determination of $\dot{G}/G$, we need also to monitor the masses of other asteroids. For this, in a separate paper (Tang and Ni, 2000), we consider all presently know 492 asteroids with diameter greater than 65 km. The results are similar.

## DISCUSSION

This simulation demonstrates that the first two goals of the ASTROD mission concept can be achieved: testing relativistic gravity with at least 3 orders of magnitude improvement and high precision measurement of solar-system parameters.

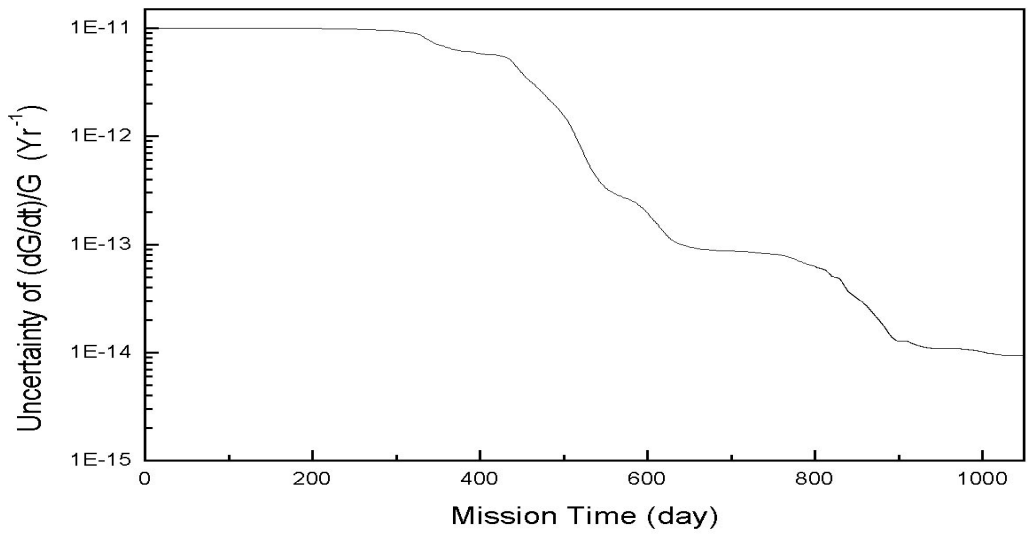

Fig. 3. Uncertainties of $\dot{G}/G$ as a function of epoch.

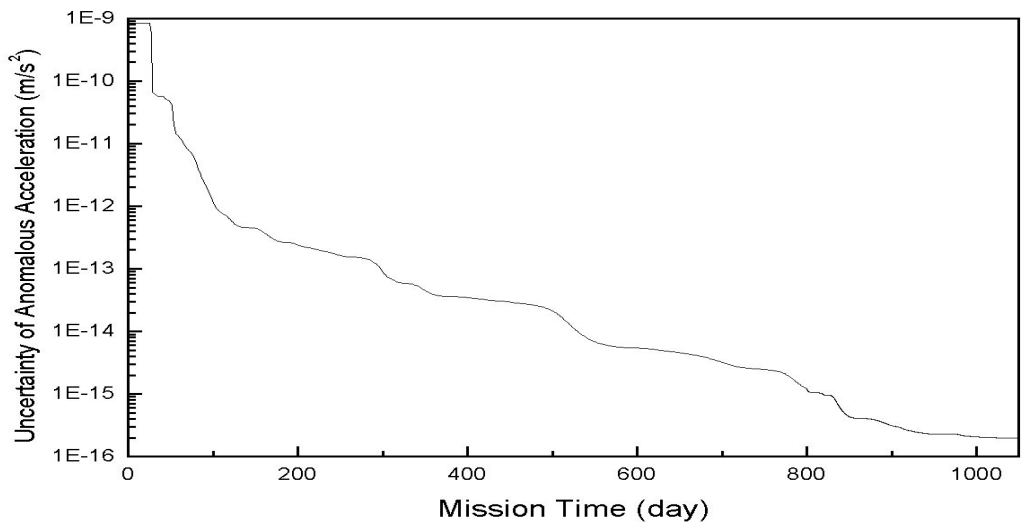

Fig. 4. Uncertainty of $a_a$ as a function of epoch.

## ACKNOWLEDGMENT

We thank the National Science Council of the Republic of China for supporting this work in part under contract Nos. NSC-88-2112-M-007-043, NSC-88-2112-M-007-045, NSC-89-2112-M-007-041, and NSC-89-2112-M-007-070.

## REFERENCES

Anderson, J. D., P. A. Laing, E. L. Lau, A. S. Liu, M. M. Nieto, and S. G. Turyshev, Indication, from Poineer 10/11, Galileo, and Ulysses Data, of an Apparent Anomalous, Weak, Long-Range Acceleration, *Phys. Rev. Lett.* **81**, 2858 (1998).

Bec-Borsenberger, A, J. Christensen-Dalsgaard, M. Cruise, A. Di Virgilio, D. Gough, M. Keiser, A. Kosovichev, C. Læmmerzahl, J. Luo, W.-T. Ni, A. Peters, E. Samain, P. H. Scherrer, J.-T. Shy, P. Touboul, K. Tsubono, A.-M.

Wu, H.-C. Yeh, Astrodynamical Space Test of Relativity using Optical Devices ASTROD--- *A Proposal Submitted to ESA in Response to Call for Mission Proposals for Two Flexi-Mission F2/F3*, January 31, 2000.

Chiou, D.-W., and W.-T. Ni, ASTROD Orbit Simulation and Accuracy of Relativistic Parameter Determination, *Advances in Space Research 25*, **6**, pp. 1259-1262 (2000).

Dickey, J. O., et al., *Science* **265**, 482 (1994).

Melliti, T., F. Fridelance, E. Samain, Study of Gravitational Theories and of the Solar Quadrupole Moment with the SORT Experiment: Solar Orbit Relativity Test, in preparation for Astronomy and Astrophysics; and references therein (2000).

MPO98 *Asteroid Viewing Guide from the Minor Planet Observer*, Bdw Publishing (1997).

Ni, W.-T., Laser Astrodynamical Mission: Concepts, Proposals and Possibilities, Plenary talk given in *the Second William Fairbank Conference on Relativistic Gravitational Experiments in Space*, 13-16 December, 1993, Hong Kong (1993).

Ni, W.-T., ASTROD mission concept and measurement of the temporal variation of the gravitation constant, pp. 309-320 in *Proceedings of the Pacific Conference on Gravitation and Cosmology, February 1-6, 1996, Seoul, Korea, ed. Y. M. Cho, C. H. Lee and S.-W. Kim* (World Scientific, Singapore, 1996).

Ni, W.-T., J.-T. Shy, S.-M. Tseng, X. Xu, H.-C. Yeh, W.-Y. Hsu, W.-L. Liu, S.-D. Tzeng, P. Fridelance, E. Samain, A.-M. Wu, Progress in Mission Concept Study and Laboratory Development for the ASTROD--- Astrodynamical Space Test of Relativity using Optical Devices, pp.105-116 in *the Proceedings of SPIE*, Vol. 3316: Small Spacecraft, Space Environments, and Instrumentation Technologies, ed. F. A. Allahdadi, E. K. Casani, and T. D. Maclay (SPIE, 1997).

Ni, W.-T., A.-M. Wu and J.-T. Shy, A Mission Concept to Measure Second-Order Relativisic Effects, Solar Angular Monentum and Low-Frequency Gravitational Waves, in *Proceedings of the Seventh Marcel Grossmann Meeting on General Relativity*, July 23-30, 1994, ed. R. T. Jantzen and G. M. Keiser, pp. 1519-1521, World Scientific (1996); and references therein.

Smith, D., Interplanetary Laser Ranging (IPLR), private communication (2000).

Standish, E. M., X. X. Newhall, J. G. Williams and W. F. Folkner, JPL Planetary and Lunar Ephemerides, DE 403/LE 403, JPL IOM 314.10-127 (1995).

Su, Z.-Y., A.-M. Wu, D. Lee, W.-T. Ni and S.-C. Lin, Asteroid Perturbations and the Possible Determination of Asteroid Masses through the ASTROD Space Mission, *Planetary and Space Science* bf, pp. 339-343 (1999).

Tang, C.-J., and W.-T. Ni, Asteroid Perturbations and Mass Determination for the ASTROD Space Mission, *presented to the 33rd COSPAR Scientific Assembly, Warsaw, 16-23 July, 2000* (2000).

Williams, J.G., X. X. Newhall, and J. O. Dickey, *Phys. Rev.* **D53**, 6730 (1996).